**Ultrafast carrier-lattice interactions and interlayer modulations of $Bi_2Se_3$ by X-ray free electron laser diffraction**

Sungwon Kim[1], Youngsam Kim[2], Jaeseung Kim[1], Sungwook Choi[1], Kyuseok Yun[1], Dongjin Kim[1], Soo Yeon Lim[1], Sunam Kim[3], Sae Hwan Chun[3], Jaeku Park[3], Intae Eom[3], Kyung Sook Kim[3], Tae-Yeong Koo[3], Yunbo Ou[4], Ferhat Katmis[4], Haidan Wen[5], Anthony Dichiara[5], Donald Walko[5], Eric C. Landahl[6], Hyeonsik Cheong[1], Eunji Sim[2], Jagadeesh Moodera[4,7], and Hyunjung Kim[1*]

[1]Department of Physics, Sogang University, Seoul, 04107, Korea

[2]Department of Chemistry, Yonsei University, Seoul, 03722, Korea

[3]Pohang Accelerator Laboratory, Pohang, 37673, Korea

[4]Francis Bitter Magnet Laboratory, Massachusetts Institute of Technology, Cambridge, MA 02139, USA

[5]Advanced Photon Source, Argonne National Laboratory, Argonne, IL 60439, USA

[6]Department of Physics, DePaul University, Chicago, IL 60614, USA

[7]Department of Physics, Massachusetts Institute of Technology, Cambridge, MA 02139, USA

Correspondence and requests for materials should be addressed to H.K. (hkim@sogang.ac.kr)

**As a 3D topological insulator, bismuth selenide ($Bi_2Se_3$) has potential applications for electrically and optically controllable magnetic and optoelectronic devices. How the carriers interact with lattice is important to understand the coupling with its topological phase. It is essential to measure with a time scale smaller than picoseconds for initial interaction. Here we use an X-ray free-electron laser to perform time-resolved diffraction to study ultrafast carrier-induced lattice contractions and interlayer modulations in $Bi_2Se_3$ thin films. The lattice contraction depends on the carrier concentration and is followed by an interlayer expansion accompanied by oscillations. Using density functional theory (DFT) and the Lifshitz model, the initial contraction can be explained by van der Waals force modulation of the confined free carrier layers. Band inversion, related to a topological phase transition, is modulated by the expansion of the interlayer distance. These results provide insight into instantaneous topological phases on ultrafast timescales.**

Bismuth selenide ($Bi_2Se_3$) is a well-known thermoelectric material[1,2], a V-VI semiconductor, and 2D van der Waals (vdW) material[3] consisting of quintuple layers (QLs). As a 3D topological insulator (TI)[4–7], $Bi_2Se_3$ has potential applications in optoelectronic[8,9] and electronic devices[4,6,10–13], and has been extensively studied. Furthermore, new applications have been suggested for heterostructures with strong spin-orbit (SOC) coupling and spin-momentum locking[14]. It is important to study ultrafast carrier-related intra- and inter-layer changes in the structure to manipulate the carriers in a topologically non-trivial, quasi-2D layered system[11–13,15–17].

Time-angle-resolved photoemission spectroscopy (trARPES) was used to study the relaxation dynamics of optically excited carriers decaying through the surface state (SS)[18], as well as oscillatory modulations of the electronic structure in the bulk and SSs[19]. Using transient laser reflections, longitudinal optical (LO) phonons with measurable carrier relaxation times were observed during electron-LO-phonon scattering[20]. These observations revealed a carrier relaxation mechanism representing a hybrid between one dominated by bulk polar phonon interactions and one dominated by surface electron-lattice interactions. How the carrier-$Bi_2Se_3$ lattice interactions couple with the vdW structures, however, needs to be investigated by direct observations. Using an X-ray free-electron laser (XFEL)[21–26], atomic-scale lattice movements and distortions can be measured directly on a timescale of a few tens of femtoseconds.

Here we describe the ultrafast time-resolved X-ray diffraction (UTXRD) experiment, carried out at PAL-XFEL[27], which studied the ultrafast carrier-induced dynamics of the $Bi_2Se_3$ lattice. The excited carriers induced lattice contraction that lasted up to 10 ps. Due to a confined layer of free carriers, high laser fluence may saturate the contraction. The contraction was followed by a lattice expansion that lasted tens of picoseconds and interlayer vibrations with breathing and interface modes. The vibrations were caused by modulating vdW forces restoring the out-of-plane distortions. The carrier-induced vdW contraction was explained by density functional theory (DFT) calculations employing the Lifshitz model. We predict that an expansion in the interlayer distance will be followed by an inversion in the topologically non-trivial band state.

Figure 1a shows the (006) Bragg reflection, accumulated over 60 shots, of a $Bi_2Se_3$ sample with 16 QLs. Additional features can be seen, such as fringes in $Q_z$ due to the sample thickness and diffuse scattering in $Q_{xy}$. The thicknesses of the sample are determined by the X-ray reflectivity and by the fringes ($\Delta Q_z$) of the (003) Bragg peak (Supplementary Fig. 1). After laser pumping, an XFEL pulse was used to probe the structure after a delay of $\Delta t$. Figure 1b shows the evolution in mean intensity of the in-plane ($\Delta Q_{xy} = 0.08 \text{ Å}^{-1}$) (006) Bragg peak from Fig. 1a. The intensity profile at $\Delta t = 0$ is plotted as a blue line. To locate the Bragg peak, the signal of the crystal truncation rods was excluded from the raw data (Supplementary Fig. 2). The peak position was taken as the centre of mass (COM). From this, the strain was calculated by the relation $\frac{\Delta Q}{Q_I}$, where $Q_I$ is the initial COM at negative $\Delta t$, and $\Delta Q$ is its change after $\Delta t$. Since no changes were observed in $Q_{xy}$, $Q_I$ and $\Delta Q$ are in $Q_z(\text{Å}^{-1})$. Note that compression is defined by a positive $\Delta Q$. The strain as a function of $\Delta t$ is shown in Figs. 1c and 1d, respectively, for a 16 QL sample with different laser fluences and for samples with different thicknesses at a fixed laser fluence of 1.1 mJcm$^{-2}$. The inset in Fig. 1c shows the strain curves when $\Delta t > 6 \text{ ps}$ for fluences of 2.2 and 3.3 mJcm$^{-2}$. After laser excitation, a contractive strain was observed. The time to reach maximum contraction depended on the sample thickness and laser fluence. Expansion was then observed, followed by relaxation with oscillations at different frequencies depending on the sample thickness.

Figures 2a and 2b show the changes in position relative to the diffraction geometry (Supplementary Fig. 3) for the (006) and (015) Bragg peaks after the laser excitation. For both peaks, the changes in $Q_z$ were almost the same; the changes in $Q_{xy}$ also showed no difference (Supplementary Fig. 4). It is possible that a change in the interlayer distance plays a dominant role in the out-of-plane dynamics.

The oscillations and relaxations are modelled by Equation (1)

$$f_{fit} = \sum_{i=1}^{m} \left[ A_{osc}^{i} \, e^{-t/\tau_{osc}^{i}} \cos 2\pi f_{osc}^{i}(t - \phi_i) \right] + \sum_{j=1}^{l} A_{rlx}^{j} \, e^{-t/\tau_{rlx}^{j}} \qquad (1)$$

where $A_{osc}$, $\tau_{osc}$, and $f_{osc}$ are, respectively, the amplitude, damping constant, and oscillation frequency for the $i$th component. $A_{rlx}$ and $\tau_{rlx}$ are, respectively, the relaxation amplitude and damping constant for the $j$th component. Figures 2c to 2f show the detailed fitting procedures and results for a 16 QL sample under 1.1 mJ/cm$^{-2}$ laser fluence. Over the entire range $\Delta t <$ 400 ps, we fit a single exponential relaxation term ($l = 1$, Fig. 2c). After subtracting this term, two superposed damping oscillations (Fig. 2d) with high ($i = 1$, Fig. 2e) and low ($i = 2$, Fig. 2f) frequencies are analysed. If the range of $\Delta t$ is extended, additional damping constants are needed to model the strain relaxation. Figures 2g and 2h show the damping constants $\tau_{rlx}^{1}$ and $\tau_{rlx}^{2}$ for both 16 and 26 QL samples, respectively; the error bars represent the 95% confidence intervals. For $\Delta t <$ 1,000 ps, the average for the 16 QL sample, $\tau_{rlx}^{1}$, is 336 ± 36 ps. However, for $\Delta t >$ 1.5 ns, when the coherent motion of the lattice diminished and thermal processes dominated, additional damping constant $\tau_{rlx}^{2}$ = 3.66 ± 0.29 ns is required. The results do not change significantly with fluence. The damping constants for the 26 QL sample are greater than those for the 16 QL one, implying that thermal diffusion takes longer through the thicker sample. Results for other sample thicknesses are given in the Supplementary Information. Figure 2i shows two distinct damping oscillation components: one with high frequency (red) and one with low frequency (blue). The measurements obtained by circularly polarized Raman spectroscopy (squares) are compared to those obtained by fitting a simple linear chain (SLC) model (line). Detailed Raman spectroscopy results are provided in the Supplementary Information. The high-frequency mode is taken as the lowest frequency interlayer breathing mode in an SLC model (Fig. 2g inset). For out of plane motions, the QLs (circles) are bonded together by vdW forces with elastic constant $K_z$. The force constant between the QLs and the substrate is taken as $K_i$. Since $K_i \ll K_z$ for the 2D materials on the substrates[28,29], the eigenmode frequencies can be written as

$$\omega_\alpha \cong \sqrt{\frac{K}{2\mu\pi^2 c^2}\left[1 - \cos\left(\frac{(\alpha-1)\pi}{N}\right)\right]} \quad (2)$$

where $\alpha = 1$ corresponds to the zero-frequency acoustic mode, and $\alpha = 2, ..., N$ to the interlayer breathing modes when $K = K_z$. The constant $c$ is the speed of light, and $\mu$ is $7.5 \times 10^{-6}$ kgm$^{-2}$ for a single QL of $Bi_2Se_3$. Fitting Equation 2 to the data in Fig. 2i gives $K_z = 5.48 \times 10^{19}$ Nm$^{-3}$ for the breathing mode. Although $\alpha = 2$ was taken as the lowest order of the breathing modes, the low-frequency mode does not resemble the breathing or shearing modes expected in interlayer vibrations. We shall call this the interface mode for the following reason. Since our UTXRD measurements are sensitive to out-of-plane motions, this mode presumably represents the sample density fluctuations caused by acoustic waves at the sample-substrate interface, subjected to strong interlayer bonding ($K_z \gg K_i$). For each sample thickness, $K_i = 4(\pi c \omega_i)^2 N\mu$. The calculated ($K_i/K_z$) ratios shown in Table 1 are comparable to those for $Bi_2Te_3$[30]. In Fig. 2j, the acoustic sound velocities calculated for the breathing mode are compared to the values obtained by Raman measurment[31] (triangles), transient laser reflection[31] (squares), and the longitudinal and transverse values (lines) calculated for the bulk[32]. Our results with a dependency on film thickness are consistent with those in Ref. 31.

This could be due to coupling of the oscillations (breathing mode) and shear components with the boundary conditions[31].

Figure 3a shows the fluence-dependent contractive strain curves ($\Delta t < 8$ ps) for the 16 QL sample described in Fig. 1c. The contractive strain data (dots) are shown with fitted error functions (lines). The maximum compression (squares) and corresponding $\Delta t$ (open circles) are also shown. With fluence ~0.4 mJcm$^{-2}$, the contraction saturates at ($\Delta Q/Q_I \sim 0.027\%$) but is tapered during the interval $\Delta t$. After contraction, the expansion reaches a maximum at ~20 ps, independent of the fluence. Figure 3b shows the maximum expansive strain as a function of fluence for the same 16 QL sample. The inset shows an enlarged range of up to 0.5 mJcm$^{-2}$. Unlike compression, the expansion is linearly related to the fluence and does not saturate.

To understand how the initial compressive strain is related to the fluence-dependent carrier density ($n$), we calculated the latter by measuring the transmittance and reflectance values (see Supplementary Information). To determine how fluence affects the contraction time, we compared our results with those obtained in a study using time-resolved angle-resolved photoemission spectroscopy (trARPES)[18] and an 800 nm fs-laser for excitation. A 26 µJcm$^{-2}$ fluence corresponds to 0.1 mJcm$^{-2}$ (24 µJcm$^{-2}$ after considering transmittance and reflectance) in our measurement. Figure 3c shows the strain curve for a fluence of 0.1 mJcm$^{-2}$. The 2.47 ps taken to reach maximum contraction is comparable to the 2.5 ps[18] taken for the carriers to completely relax by scattering from the higher-lying states, after they populated the SS and bulk conduction band (BCB) within 0.7 ps of excitation. Since the X-ray penetrated the entire 16 QLs, the SS population is negligible. The observed contractions are therefore attributed to the carriers that populated the BCB.

To relate the density of the excited carriers to the magnitude of the lattice contraction, we used the Lifshitz model[33]. By considering the dielectric function, this model explains the vdW force between metallic surfaces separated by distance ($l$), which is small compared to the wavelengths of the fluctuating field. We assume that the excited carriers formed a charge slab inside the QL and each slab is separated by the distance $l$. At the high carrier density limit, where the plasma frequency is much larger than the Drude damping term in the dielectric function, the Drude model gives the effective pressure between charged slabs[26] as

$$\Delta F = \frac{\xi \hbar e \sqrt{n}}{32\pi^2 \sqrt{2m\epsilon_0} l^3} \quad (3)$$

where $\hbar$ is the Planck constant, $e$ is the elementary electron charge, $\epsilon_0$ is the vacuum permittivity, $n$ is the carrier density, and $\xi = 2.04$[26] is the Matsubara frequency. The stress is calculated by multiplying the elastic tensor component along the c-axis ($C_{33}$) by the maximum contraction in Fig. 3a. Figure 4a shows the compressive stress on the 16 QL sample (circle) as a function of $n$ and the fit with Equation (3) (line). The distance $l$ is 3.94 Å from the fit, larger than the interlayer distance of $Bi_2Se_3$ (2.58 Å) calculated by DFT. This implies that in $Bi_2Se_3$, the charge density (CD) within a QL is responsible for the vdW attraction. Note that the data with $n > 3.564 \times 10^{20}$ were excluded from the fit due to saturation (inset of Fig. 4a).

After fitting Equation (3) to the data, we wanted to explore the meaning of parameter $l$ in a QL. Using DFT, we calculated the CD of the valence band (VB) and the conduction band (CB). The calculations are described in the Methods section. Figure 4b shows the CD and atomic configuration in a single QL along the c-axis direction. DFT geometry optimization gave the following intra-layer distances: 1.58 Å for Se(outer)–Bi, 1.90 Å for Bi–Se(inner), and 2.58 Å for the interlayer distance ($s$), which is the distance between Se(outer)-Se(outer) layers. The black dashed lines indicate indentations $l/2$ from each edge of the QL (9.55Å). The CD of the VB (cyan) shows electrons distributed around the Se atoms. The CD of the CB (purple) has a distribution that is diffuse within the dashed lines and decreases sharply outwards, reaching a minimum within the interlayer distance. As the region between the dashed lines contains most of the CB carriers (> 70%), we consider it as a charged slab with thickness $d = 9.55 - l$ (Å). In other words, the carriers confined within the slab ($d$) are responsible for the compressive strain.

To investigate how interlayer distance modulation affects the topological phase of $Bi_2Se_3$[16,34], we calculated the band structures for different interlayer distances ($s + \Delta s$) while keeping the atomic structure within the QL fixed. The band structure near the $\Gamma$ point and corresponding projected density of states (PDOS) with interlayer distance changes ($\Delta s$ in inset) are overlaid in Figure 4c. The atomic orbitals are indicated by the coloured lines. The PDOS shows that as $\Delta s$ expands, $p_z$ orbitals contribute to the inversion state, but the $p_z$ orbital contribution remains constant regardless of inversion. According to our observations, with increasing laser fluence, the lattice expands linearly after the carrier-induced contraction. For a 3.3 mJcm$^{-2}$ fluence, it expands to 29.43 Å, 0.79 Å larger than during the equilibrium state. We attribute most of the expansion to a coherent interlayer distance modulation since neither significant in-plane lattice change nor evidence of lattice thermalization has been observed on this time scale. In the DFT calculations, the energy gap between the inverted states narrows when the interlayer distance expands. The band is inverted when the interlayer distance expansion exceeds 0.34 Å, predicting a topological phase transition to a normal insulator. In this work, the largest observed interlayer distance expansion is 0.26 Å; the corresponding energy band gap is 0.06 eV — a reduction of 0.24 eV compared to the equilibrium state. This suggests that the topological phase is optically tunable on an ultrafast timescale.

In conclusion, we have studied the lattice dynamics of $Bi_2Se_3$ on an ultrafast timescale, observing the consecutive dynamical phases induced by carrier excitation. During $\Delta t < \sim 6\ ps$, due to the presence of a concentrated layer of carriers within a QL, an impulsive contraction is induced by modulating vdW forces. By performing DFT calculations with the Lifshitz model, we found that the carrier layer is 5.61 Å thick, with 70% of the carriers occupying the CB. As the contraction is reversed by carrier-lattice interaction, we observed breathing and interface mode oscillations. The vibration modes of the interlayer oscillations are assigned by an SLC model. The linear relationship between the sample thickness and damping constants of the oscillations and relaxations suggests a role for thermal diffusion. The $Bi_2Se_3$ samples undergo maximum interlayer expansion at the beginning of the interlayer oscillations. How the topologically inverted states near the Fermi level are modulated is explained quantitatively by the expansion of the interlayer distance. With an expansion of 0.26 Å, the bulk band gap is expected to narrow to 0.06 eV. Hence a topological phase transition to a normal insulator is predicted for an expansion larger than 0.34 Å. These findings may provide insights relevant to applications that utilize the ultrafast topological states in 2D TIs.

**Methods**

**Sample preparation**

The $Bi_2Se_3$ films were grown on insulating $Al_2O_3$ (0001) substrates, along with the [001], using a molecular beam epitaxy (MBE) co-evaporation method. The ultra-high vacuum (UHV) system has a base pressure below $3.75 \times 10^{-10}$ Torr. Before film growth, the substrate was cleaned with acetone and isopropyl alcohol and dried by dry nitrogen. The substrate was then loaded into our UHV system and degassed at 600°C for 10 min, and then at 800°C for 30 min. High-purity Bi (99.999%) and Se (99.999%) were co-evaporated from Knudsen cells adjusted to obtain a 2:3 compositional ratio in the grown film. Growth rates were measured by two independent *in-situ* quartz crystal monitors. The $Bi_2Se_3$ films were grown at a substrate temperature of 240°C under Se-rich conditions, with a typical Se/Bi flux ratio of 10. The typical growth rate was 1 QL min$^{-1}$ (QL, where 1 QL ≈ 0.96 nm). A protective 5-nm amorphous $Al_2O_3$ layer was then deposited *in-situ* at room temperature.

**X-ray reflectivity measurements**

X-ray reflectivity measurements were taken with a D8 Discover diffractometer (Bruker Corporation, Billerica, MA, USA). An 8.05 keV (Cu $K_\alpha$) X-ray beam was collimated by a Göbel mirror and then filtered by a Ni filter to exclude the $K_\beta$ emission.

**Ultrafast time-resolved X-ray diffraction**

With a 500-fs time resolution and delays up to hundreds of ps, the UTXRD measurements were carried out at the X-ray Scattering and Spectroscopy (XSS) beamline at PAL-XFEL (Pohang, Korea)[35]. A monochromatized X-ray pulse with a width of ~20 fs was used to probe the sample $\Delta t$ after the optical laser pump. The bandwidth was $\Delta E/E \sim 1.6 \times 10^{-4}$ at 9.7 keV. The monochromator was a Si(111) double crystal. The X-ray pulses were focused by a series of beryllium compound refractive lenses (CRL) to a spot of size $20 \times 20$ μm$^2$ full width at half maximum (FWHM) at the sample location. A Ti:sapphire laser with 800 nm wavelength and pulse width of 100 fs FWHM was used as the optical pump. It was p-polarized to the sample surface, and its angle of incidence was 9.6° larger than the X-ray. The laser gave effective fluences between 0.1 and 3.3 mJcm$^{-2}$ on the sample surface. After a time delay of $\Delta t$, measurements were taken with the laser on and off, and then compared. The X-ray scattering patterns were recorded by a 2D "multiport charged-coupled device" (MPCCD) with a 10–30 Hz frequency.

Time-resolved X-ray diffraction measurements with delays of up to 8 ns were taken at the 7-ID-C beamline of the Advanced Photon Source (USA). A Ti:sapphire laser with 50 fs FWHM pulse duration, 800 nm wavelength, and 1 kHz repeating frequency was used as the optical pump. The sample was probed by 10 keV X-ray pulses with an FWHM duration of 90 ps. Collinear with the X-ray beam, the pump laser gave fluences between 0.7 and 8.8 mJcm$^{-2}$ at the sample surface. A 2D Pilatus detector was used 1 m from the sample.

**Density functional theory (DFT) calculations**

The electronic structure of Bi$_2$Se$_3$ was calculated using DFT at the level of the generalized gradient approximation (GGA), in particular, Perdew-Burke-Ernzerhof (PBE)[36] functional including spin-orbit coupling (SOC) implemented in the Vienna Ab Initio Simulation Package (VASP)[37]. The projector augmented wave (PAW) method[38] was used in the calculations.

Specifically, 5*s*, 6*s*, 5*p*, 6*p*, and 5*d* of bismuth, and 4*s* and 4*p* of selenium, were included as the valence states. The convergence criterion for the self-consistency calculation was $10^{-6}$ eV, and the kinetic energy cut-off was 340 eV with accurate precision mode. The Brillouin zone was sampled using $13 \times 13 \times 13$ k-points according to the Monkhorst-Pack scheme. The tetrahedron method with Blöchl corrections was used, except for the band structure calculations.

To optimize the geometry, we fixed the lattice constants of a hexagonal unit cell and released the atomic positions. The in-plane lattice parameters "a" and "c" were set to 4.17 Å[17] and 28.64 Å[39], respectively. The optimization procedure continued until the Hellmann-Feynman forces became smaller than 0.01 eV/Å, and the vdW interactions between the QLs could be taken into account by Grimme's vdW correction (D2)[40].

The CD distribution in the z-direction (Fig. 4b) was obtained based on the local integral curve using Multiwfn[41]. The curve is defined as $I_L(z) = \int \int p(x, y, z) \, dxdy$. We summed the charge densities of the carriers in the VB and CB near the $\Gamma$ point. The k-points contained 1/4 of the $\Gamma \to M$, $\Gamma \to K$ and $\Gamma \to A$ paths.

The band structure (Fig. 4c) was calculated along the symmetry points (K, $\Gamma$, M). With Gaussian smearing (0.01 eV broadening), each line contained 80 such points. Various structures in the interlayer distance were calculated while keeping the atomic structure within a QL fixed. We obtained the PDOS at the $\Gamma$ point by incorporating the *lm*-decomposed scheme into the band structure calculation.

# References

1. Mishra, S. K., Satpathy, S. & Jepsen, O. Electronic structure and thermoelectric properties of bismuth telluride and bismuth selenide. *J. Phys. Condens. Matter* **9**, 461–470 (1997).

2. Ghaemi, P., Mong, R. S. K. & Moore, J. E. In-plane transport and enhanced thermoelectric performance in thin films of the topological insulators $Bi_2Te_3$ and $Bi_2Se_3$. *Phys. Rev. Lett.* **105**, 166603 (2010).

3. Geim, A. K. & Grigorieva, I. V. Van der Waals heterostructures. *Nature* **499**, 419–425 (2013).

4. Zhang, H. *et al.* Topological insulators in $Bi_2Se_3$, $Bi_2Te_3$ and $Sb_2Te_3$ with a single Dirac cone on the surface. *Nat. Phys.* **5**, 438–442 (2009).

5. Moore, J. E. The birth of topological insulators. *Nature* **464**, 194–198 (2010).

6. Qi, X. L. & Zhang, S. C. Topological insulators and superconductors. *Rev. Mod. Phys.* **83**, 1057-1110 (2011).

7. Fu, L., Kane, C. L. & Mele, E. J. Topological insulators in three dimensions. *Phys. Rev. Lett.* **98**, 106803 (2007).

8. Zhang, H., Zhang, X., Liu, C., Lee, S. T. & Jie, J. High-Responsivity, High-Detectivity, Ultrafast Topological Insulator $Bi_2Se_3$/Silicon Heterostructure Broadband Photodetectors. *ACS Nano* **10**, 5113–5122 (2016).

9. Wang, F. *et al.* Submillimeter 2D $Bi_2Se_3$ Flakes toward High-Performance Infrared Photodetection at Optical Communication Wavelength. *Adv. Funct. Mater.* **28**,

1802707 (2018).

10. Mellnik, A. R. *et al.* Spin-transfer torque generated by a topological insulator. *Nature* **511**, 449–451 (2014).

11. Luo, L. *et al.* Ultrafast manipulation of topologically enhanced surface transport driven by mid-infrared and terahertz pulses in $Bi_2Se_3$. *Nat. Commun.* **10**, 607 (2019).

12. Checkelsky, J. G., Hor, Y. S., Cava, R. J. & Ong, N. P. Bulk band gap and surface state conduction observed in voltage-tuned crystals of the topological insulator $Bi_2Se_3$. *Phys. Rev. Lett.* **106**, 196801 (2011).

13. Kim, D. *et al.* Surface conduction of topological Dirac electrons in bulk insulating $Bi_2Se_3$. *Nat. Phys.* **8**, 459–463 (2012).

14. Katmis, F. *et al.* A high-temperature ferromagnetic topological insulating phase by proximity coupling. *Nature* **533**, 513–516 (2016).

15. Flötotto, D. *et al.* In Situ Strain Tuning of the Dirac Surface States in $Bi_2Se_3$ Films. *Nano Lett.* **18**, 5628–5632 (2018).

16. Aramberri, H. & Muñoz, M. C. Strain-driven tunable topological states in $Bi_2Se_3$. *J. Phys.: Mater.* **1**, 015009 (2018).

17. Hsieh, D. *et al.* A tunable topological insulator in the spin helical Dirac transport regime. *Nature* **460**, 1101–1105 (2009).

18. Sobota, J. A. *et al.* Ultrafast optical excitation of a persistent surface-state population in the topological insulator $Bi_2Se_3$. *Phys. Rev. Lett.* **108**, 117403 (2012).

19. Sobota, J. A. *et al.* Distinguishing bulk and surface electron-phonon coupling in the

topological insulator Bi$_2$Se$_3$ using time-resolved photoemission spectroscopy. *Phys. Rev. Lett.* **113**, 117401 (2014).

20. Glinka, Y. D. *et al.* Ultrafast carrier dynamics in thin-films of the topological insulator Bi$_2$Se$_3$. *Appl. Phys. Lett.* **103**, 151903 (2013).

21. Trigo, M. *et al.* Fourier-transform inelastic X-ray scattering from time- and momentum-dependent phonon-phonon correlations. *Nat. Phys.* **9**, 790–794 (2013).

22. Zhu, D. *et al.* Phonon spectroscopy with sub-meV resolution by femtosecond x-ray diffuse scattering. *Phys. Rev. B* **92**, 054303-1 (2015).

23. Fritz, D. M. Ultrafast bond softening in Bismuth : *Science* **315**, 633–637 (2007).

24. Zalden, P. *et al.* Femtosecond x-ray diffraction reveals a liquid–liquid phase transition in phase-change materials. *Science* **364**, 1062–1067 (2019).

25. Lindenberg, A. M., Johnson, S. L. & Reis, D. A. Visualization of atomic-scale motions in materials via femtosecond X-ray scattering techniques. *Annu. Rev. Mater. Res.* **47**, 425–449 (2017).

26. Mannebach, E. M. *et al.* Dynamic optical tuning of interlayer interactions in the transition metal dichalcogenides. *Nano Lett.* **17**, 7761–7766 (2017).

27. Kang, H. S. *et al.* Hard X-ray free-electron laser with femtosecond-scale timing jitter. *Nat. Photonics* **11**, 708–713 (2017).

28. Tan, P. H. *et al.* The shear mode of multilayer graphene. *Nat. Mater.* **11**, 294–300 (2012).

29. Zhao, Y. *et al.* Interlayer breathing and shear modes in few-trilayer MoS$_2$ and WSe$_2$.

*Nano Lett.* **13**, 1007–1015 (2013).

30. Zhao, Y. *et al.* Interlayer vibrational modes in few-quintuple-layer $Bi_2Te_3$ and $Bi_2Se_3$ two-dimensional crystals: Raman spectroscopy and first-principles studies. *Phys. Rev. B* **90**, 245428 (2014).

31. Glinka, Y. D., Babakiray, S., Johnson, T. A., Holcomb, M. B. & Lederman, D. Acoustic phonon dynamics in thin-films of the topological insulator $Bi_2Se_3$. *J. Appl. Phys.* **117**, 165703 (2015).

32. Giraud, S., Kundu, A. & Egger, R. Electron-phonon scattering in topological insulator thin films. *Phys. Rev. B.* **85**, 035441 (2012).

33. Lifshitz, E. M. The theory of molecular attractive forces between solids. *Perspect. Theor. Phys.* **2**, 329–349 (1992).

34. Yang, W. J. *et al.* Tuning of topological Dirac states via modification of van der Waals gap in strained ultrathin $Bi_2Se_3$ films. *J. Phys. Chem. C* **122**, 23739–23748 (2018).

35. Park, J. *et al.* Design of a hard X-ray beamline and end-station for pump and probe experiments at Pohang Accelerator Laboratory X-ray Free Electron Laser facility. *Nucl. Instruments Methods Phys. Res. Sect. A Accel. Spectrometers, Detect. Assoc. Equip.* **810**, 74–79 (2016).

36. Kresse, G. & Furthmüller, J. Efficient iterative schemes for ab initio total-energy calculations using a plane-wave basis set. *Phys. Rev. B* **54**, 11169–11186 (1996).

37. Perdew, J. P., Burke, K. & Ernzerhof, M. Generalized gradient approximation made simple. *Phys. Rev. Lett.* **77**, 3865–3868 (1996).

38. Blöchl, P. E. Projector augmented-wave method. *Phys. Rev. B* **50**, 17953–17979 (1994).

39. Lind, H., Lidin, S. & Häussermann, U. Structure and bonding properties of $(Bi_2Se_3)_m(Bi_2)_n$ stacks by first-principles density functional theory. *Phys. Rev. B - Condens. Matter Mater. Phys.* **72**, 184101 (2005).

40. Allouche, A. Semiempirical GGA-Type Density Functional Constructed with a Long-Range Dispersion Correction. *J. Comput. Chem.* **27**, 1787–1799 (2006).

41. Lu, T. & Chen, F. Multiwfn: A multifunctional wavefunction analyzer. *J. Comput. Chem.* **33**, 580–592 (2012).

**ACKNOWLEDGEMENTS**

We thank Sanghoon Song and Aymeric Robert for fruitful discussions. This research was supported by the National Research Foundation of Korea (NRF-2015R1A5A1009962, 2019R1A6B2A02100883). Y.K. and E.S. acknowledge the support from NRF-2020R1A2C2007468. Use of the Advanced Photon Source was supported by the Office of Basic Energy Science, under the Office of Science of the US Department of Energy (Contract No. DE-AC02-06CH11357). The experiments were carried out at the XSS beamline of PAL-XFEL (experiment no. 2017-1st-FXS-006 and 2018-1st-XSS-010) funded by the Ministry of Science and ICT of Korea. The work at MIT was supported by the Center for Integrated Quantum Materials (NSF-DMR 1231319), the NSF Grant DMR 1700137, NSF CONVERGENCE ACCELERATOR TRACK C: SYN OIA-2040620, and ONR Grants N00014-16-1-2657 and N00014-20-1-2306.

**Corresponding author**

Correspondence to: Hyunjung Kim

**Author contributions**

H.K. supervised and coordinated all aspects of the project. The UTXRD measurements at PAL-XFEL were carried out by S.K., J.K., S.C., K.Y., D.K., S.K., S.H.C., J.P., I.E., K.K, T-Y.K., and H.K. The UTXRD measurements at APS were carried out by S.K., K.Y., H.W., A.D., D.W., E.C.L., and H.K. The UTXRD data were analysed by S.K. The $Bi_2Se_3$ thin films were grown by Y.O. and F.K. under the supervision of J.S.M. The XRR measurements were carried out by J.K. and S.K. The theoretical analyses and density functional theory calculations were carried out by S.K. and Y.K., under the supervision of E.S. and H.K. S.Y.L. carried out the Raman spectroscopic measurements under the supervision of H.C. S.K., Y.K., E.S., and H.K. wrote the paper. All authors discussed the results and commented on the manuscript.

**Figure 1. Strain evolution observed at (006) Bragg peak.** (a) The (006) Bragg reflection before laser excitation. (b) In-plane ($\Delta Q_{xy}$ = 0.08 Å$^{-1}$)-averaged (006) Bragg peak after the time delay. The intensity profile at $\Delta t$ = 0 is marked by the blue line. (c) Strain evolution in a 16 QL Bi$_2$Se$_3$ film under different laser fluences. Contractive strains were observed between time delays of 0 and 6 ps. The lattice was subsequently released, and the strain rapidly turned expansive and was accompanied by oscillations. (d) Strain evolution in Bi$_2$Se$_3$ films with 7, 16, 26, 43, and 78 QLs, with fluence of 1.1 mJcm$^{-2}$.

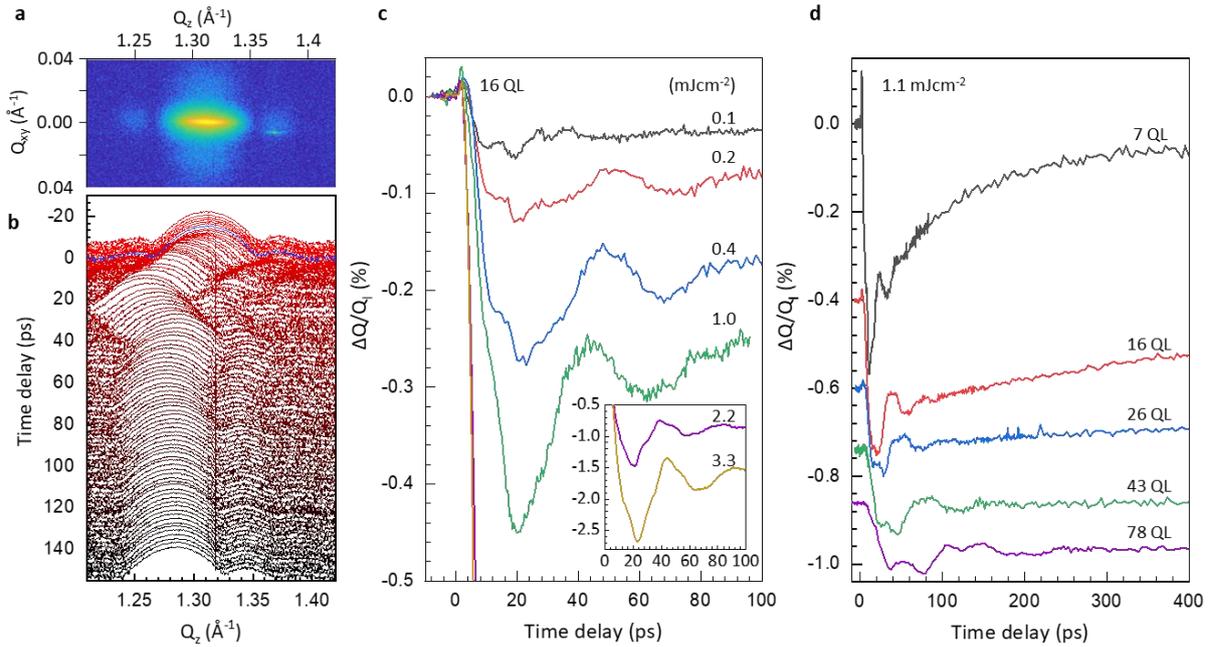

**Figure 2. Interlayer vibration modes and relaxation fits to strain curves.** (a–b) Position change of the (006) and (015) Bragg peaks in the $Q_z$ and $Q_{zy}$ directions. (c–f) Example strain curve fitting example with a 16 QL sample and 1.1 mJcm$^{-2}$ fluence. (g) The relaxation damping constant $\tau^1_{rlx}$ for 16QL (squares) and 26 QL (circles). The red and blue symbols correspond to the values derived from data obtained at XFEL and the synchrotron source, respectively. (h) $\tau^2_{rlx}$ for 16QL (squares) and 26 QL (circles) derived from data obtained at the synchrotron. (i) The oscillation frequencies for the breathing (red) and interface (blue) modes, for Bi$_2$Se$_3$ films with different thicknesses. The error bars denote the 95% confidence intervals. The measured Raman frequencies are plotted as squares. The solid blue line represents the fit to the breathing mode model. (j) The propagation velocities for the breathing mode are plotted as blue circles. The error bars denote the 95% confidence intervals. The triangle and square represent the measurements obtained by Raman spectroscopy (Ref. 30) and ultrafast laser reflection (Ref. 31), respectively.

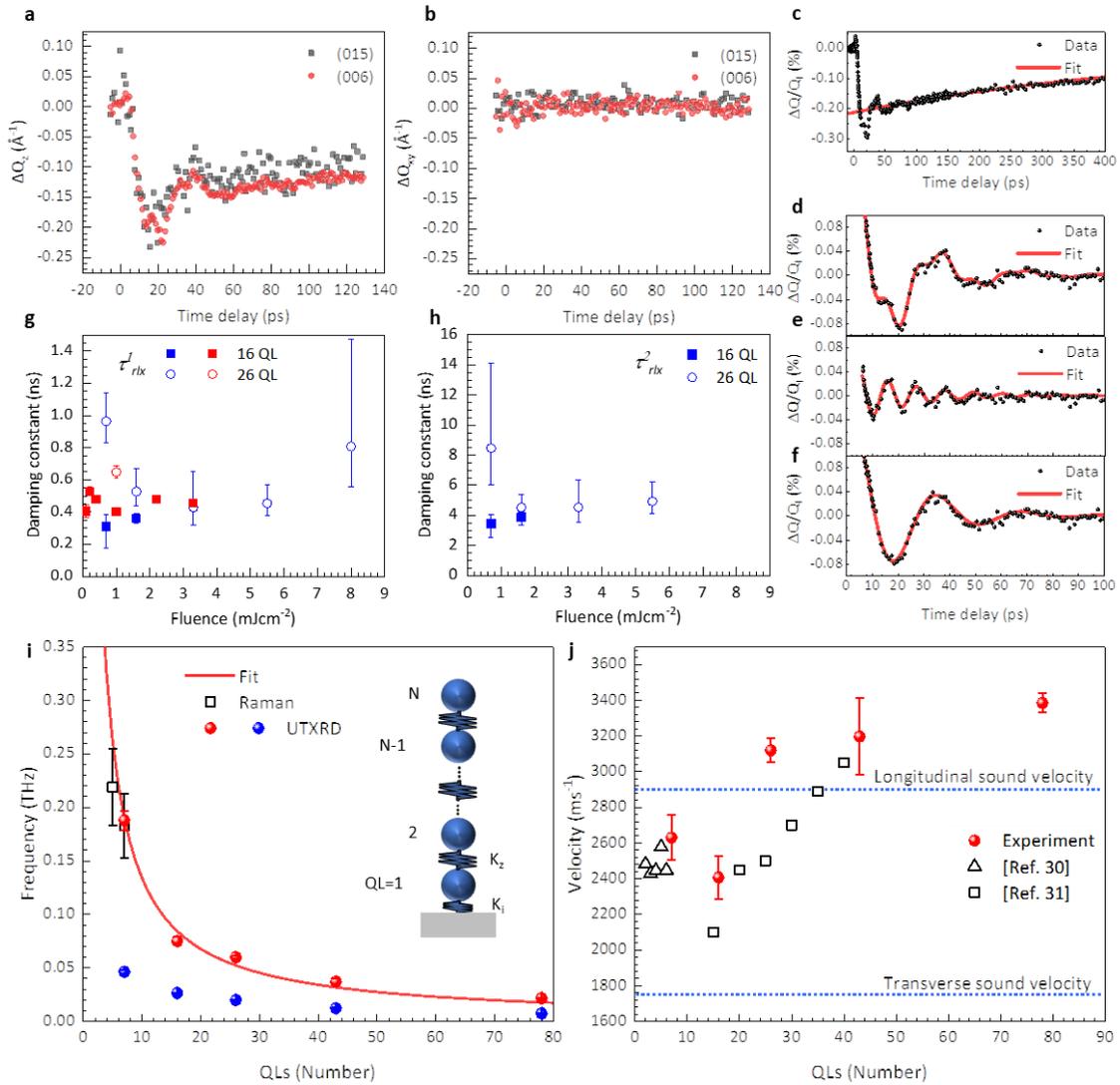

**Figure 3. The interlayer contractive strain and maximum expansive strain on a 16 QL Bi$_2$Se$_3$ sample.** (a) Contractive strain for different fluences up to time delay of 8 ps. The data points (dots) are plotted with the error functions (lines). The maximum contractive strains and their corresponding time delays are plotted as squares and open circles, respectively. (b) The maximum expansive strain as a function of laser fluence for the 16 QL sample from Fig. 1d. The laser fluence ranges from 0.1 to 3.3 mJcm$^{-2}$. The maximum lattice expansion scales linearly with fluence. The maximum expansive strains were taken as the minima in the strain curves. For each time delay, the centre of mass is averaged over the measurements, and the error bars represent the standard deviation. (c) Compressive strain as a function of time delay for a 1 mJcm$^{-2}$ fluence. The maximum contraction occurred at a time delay of 2.47 ps.

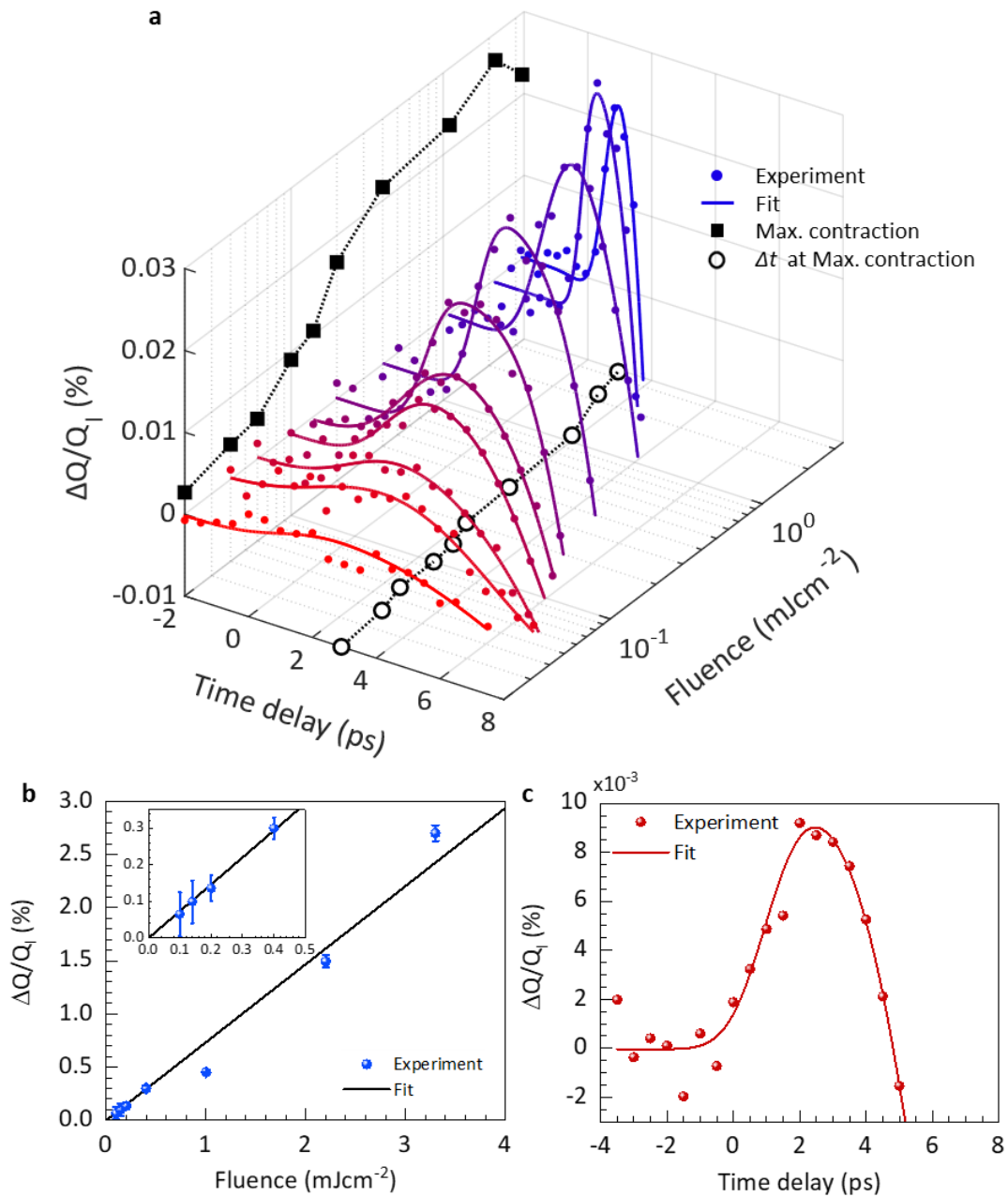

**Figure 4. Theoretical analysis of interlayer distance contraction and expansion.** (a) The maximum compressive stress is plotted against the carrier density extracted from Fig. 3a (red circle), with the Lifshitz model (line) overlaid. The stress data for carrier densities over 3.564 × $10^{20}$ (with 0.4 mJcm$^{-2}$ fluence) were excluded since they were in the saturation regime. The inset shows the entire dataset. (b) The calculated charge density (CD) distribution along the c-axis of the crystal and the atomic configuration in a QL. To display the carrier distribution, we have aligned the y-axis of the CD distribution with the atomic positions. The dashed lines represent the boundaries of the metallic surface in the Lifshitz model. (c) The projected density of states (PDOS) of modulated interlayer distance, *Δs*, and the corresponding band structure in the vicinity of the Γ point. The coloured lines indicate contributions by atomic orbitals. Band inversion occurs as the interlayer distance expands.

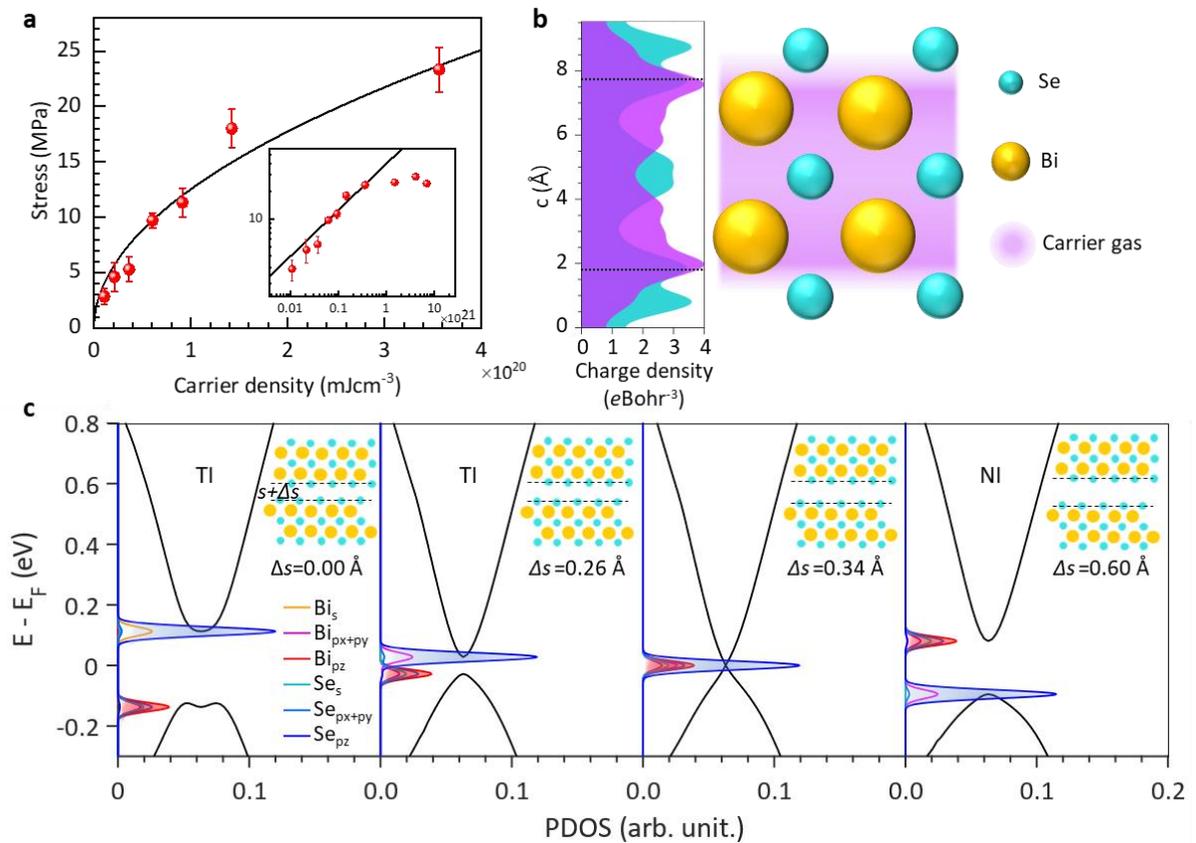

**Table 1.** $K_i$ values obtained from the results with a simple linear chain model with a nonzero substrate force constant and $K_i/K_z$ ratios.

| Thickness (QLs) | 7 | 16 | 26 | 43 | 78 |
|---|---|---|---|---|---|
| $K_i$ (×10$^{18}$ Nm$^{-3}$) | 4.35 | 3.26 | 2.98 | 1.90 | 1.24 |
| $K_i / K_z$ (×10$^{-2}$) | 8.43 | 7.67 | 4.18 | 2.54 | 1.48 |

Supplementary information

# Ultrafast carrier-lattice interactions and interlayer modulations of Bi$_2$Se$_3$ by X-ray free electron laser diffraction

Sungwon Kim[1], Youngsam Kim[2], Jaeseung Kim[1], Sungwook Choi[1], Kyuseok Yun[1], Dongjin Kim[1], Soo Yeon Lim[1], Sunam Kim[3], Sae Hwan Chun[3], Jaeku Park[3], Intae Eom[3], Kyung Sook Kim[3], Tae-Yeong Koo[3], Yunbo Ou[4], Ferhat Katmis[4], Haidan Wen[5], Anthony Dichiara[5], Donald Walko[5], Eric C. Landahl[6], Hyeonsik Cheong[1], Eunji Sim[2], Jagadeesh Moodera[4,7], and Hyunjung Kim[1*]

[1]Department of Physics, Sogang University, Seoul, 04107, Korea

[2]Department of Chemistry, Yonsei University, Seoul, 03722, Korea

[3]Pohang Accelerator Laboratory, Pohang, 37673, Korea

[4]Francis Bitter Magnet Laboratory, MIT, Cambridge, MA 02139, USA

[5]Advanced Photon Source, Argonne National Laboratory, Argonne, IL 60439, USA

[6]Department of Physics, DePaul University, Chicago, IL 60614, USA

[7]Department of Physics, MIT, Cambridge, MA 02139, USA

**This file includes**

    Supplementary Notes 1-4

    Supplementary Figures 1-10

    Supplementary Tables 1-3

    Supplementary References

**Supplementary Notes**

**Supplementary Note 1. Film thickness determination by X-ray reflectivity measurements**

We measured X-ray reflectivity (XRR) and X-ray diffraction (XRD) to determine the thickness of the samples. Supplementary Fig. 1 shows (a) XRR results and fitting for the 16, 26, 43, and 78 QL samples and (b) (006) Bragg peak and fringes from 5, 7, and 16 QL samples. In XRR measurement, the wave vector transfer $Q_z$ in Supplementary Fig. 1a is $4\pi/\lambda \sin\theta$, where $\theta$ is the incident angle, and $\lambda$ is the X-ray wavelength. Kiessig fringes from the $Bi_2Se_3$ surface and the interface between the substrate in the reflectivity curves are from the thickness $Bi_2Se_3$. The recursive Parratt formalism with modified Fresnel coefficients was used to fit the XRR data[1]. The electron density profile of $Bi_2Se_3$ and $Al_2O_3$ was obtained with the surface and interface of tanh-profile for the roughness distribution function. The layer thicknesses are obtained as 157.0, 256.0, 430.3, and 780.3 Å from the Fourier transform of the XRR result.[2]

Sample thicknesses of 5, 7, and 16 QL are obtained from the XRD measurements. The thickness ($d$) is calculated from the fringe spacing ($\Delta Q_z$) of (006) Bragg peak with the relation $d = \frac{2\pi}{\Delta Q_z}$. They are 47.0, 66.1, and 161.1Å.

**Supplementary Note 2. Reduction of crystal truncation rods**

Crystal truncation rods (CTRs) from multiple domains are detected at the (006) Bragg reflection. Supplementary Fig S2 shows the mean valued intensity profile along in-plane $Q_{xy}$ of Fig. 1a in the main text. The blue line is obtained from raw data. The sharp spikes from the CTR peak are clearly seen. Crystal truncation rods are excluded from the raw data to trace the Bragg peak position. The intensity profile shows only the central Bragg peak and fringes after excluding the column pixels of the detector image, which contains CTRs (red).

**Supplementary Note 3. Raman measurement of Bi₂Se₃**

Hexagonally close-packed Bi₂Se₃ with D$_{3d}$ point group has the irreducible representations of $\Gamma_{bulk} = 3E_u + 3A_{2u} + 2E_g + 2A_{1g}$ , $\Gamma_{odd} = \frac{5N-1}{2}(A_{1g} + E_g) + \frac{5N+1}{2}(A_{2u} + E_u)$ (N=1,3,5…), and $\Gamma_{even} = \frac{5N}{2}(A_{1g} + A_{2u} + E_g + E_u)$ ($N = 2,4,6…$) , for bulk, odd-, and even-layer, respectively. In order to separate breathing modes ($A_g$ modes) and shear modes ($E_g$ modes), we used a λ/4 waveplate to convert the linearly polarized light to circularly polarized light, or *vice versa*. Incident and scattered light propagate along the z-axis of samples in back-scattering geometry. Raman intensity is proportional to $|\langle\sigma_s|R|\sigma_i\rangle|^2$, where R represents the Raman tensor of each mode and $\sigma_{s,i}$ represents a vector of the circularly polarized incident and scattered light with $\sigma\pm = \frac{1}{\sqrt{2}}(1 \mp i \ 0)$ in back-scattering geometry. Raman tensors of A$_{1g}$ and E$_g$ mode are given by $R(A_{1g}) = \begin{pmatrix} a & 0 & 0 \\ 0 & a & 0 \\ 0 & 0 & b \end{pmatrix}$, $R(E_{g,1}) = \begin{pmatrix} c & 0 & 0 \\ 0 & -c & d \\ 0 & d & 0 \end{pmatrix}$, and $R(E_{g,2}) = \begin{pmatrix} 0 & -c & -d \\ -c & 0 & 0 \\ -d & 0 & 0 \end{pmatrix}$, respectively.

For the same circular polarization [(σ+σ+) or (σ-σ-)], the intensity of $A_{1g}$ mode is proportional to a² whereas the intensity of $E_g$ mode is zero.

$$|\langle\sigma+|R|\sigma+\rangle|^2_{A_{1g}} \propto \frac{1}{4}\left|(1 \ i \ 0)\begin{pmatrix} a & 0 & 0 \\ 0 & a & 0 \\ 0 & 0 & b \end{pmatrix}\begin{pmatrix} 1 \\ -i \\ 0 \end{pmatrix}\right|^2 = \frac{1}{4}|a+a+0|^2 = a^2$$

$$\left|\langle\sigma-|R|\sigma-\rangle\right|^2_{A_{1g}} \propto \frac{1}{4}\left|(1\ -i\ 0)\begin{pmatrix}a & 0 & 0\\ 0 & a & 0\\ 0 & 0 & b\end{pmatrix}\begin{pmatrix}1\\ i\\ 0\end{pmatrix}\right|^2 = \frac{1}{4}|a+a+0|^2 = a^2$$

$$\left|\langle\sigma+|R|\sigma+\rangle\right|^2_{E_{g,1}} \propto \frac{1}{4}\left|(1\ i\ 0)\begin{pmatrix}c & 0 & 0\\ 0 & -c & d\\ 0 & d & 0\end{pmatrix}\begin{pmatrix}1\\ -i\\ 0\end{pmatrix}\right|^2 = \frac{1}{4}|c-c|^2 = 0$$

$$\left|\langle\sigma-|R|\sigma-\rangle\right|^2_{E_{g,1}} \propto \frac{1}{4}\left|(1\ -i\ 0)\begin{pmatrix}c & 0 & 0\\ 0 & -c & d\\ 0 & d & 0\end{pmatrix}\begin{pmatrix}1\\ i\\ 0\end{pmatrix}\right|^2 = \frac{1}{4}|c-c|^2 = 0$$

$$\left|\langle\sigma+|R|\sigma+\rangle\right|^2_{E_{g,2}} \propto \frac{1}{4}\left|(1\ i\ 0)\begin{pmatrix}0 & -c & -d\\ -c & 0 & 0\\ -d & 0 & 0\end{pmatrix}\begin{pmatrix}1\\ -i\\ 0\end{pmatrix}\right|^2 = \frac{1}{4}|ci-ci|^2 = 0$$

$$\left|\langle\sigma-|R|\sigma-\rangle\right|^2_{E_{g,2}} \propto \frac{1}{4}\left|(1\ -i\ 0)\begin{pmatrix}0 & -c & -d\\ -c & 0 & 0\\ -d & 0 & 0\end{pmatrix}\begin{pmatrix}1\\ i\\ 0\end{pmatrix}\right|^2 = \frac{1}{4}|-ci+ci|^2 = 0$$

On the other hand, the intensity of $A_{1g}$ mode is zero for the opposite circular polarization [(σ+σ-) or (σ-σ+)] whereas the intensity of $E_g$ mode is proportional to $c^2$. Since two $E_g$ modes are degenerate, the polarization dependence has a superposed form of the two modes.

$$\left|\langle\sigma+|R|\sigma-\rangle\right|^2_{A_{1g}} \propto \frac{1}{4}\left|(1\ i\ 0)\begin{pmatrix}a & 0 & 0\\ 0 & a & 0\\ 0 & 0 & b\end{pmatrix}\begin{pmatrix}1\\ i\\ 0\end{pmatrix}\right|^2 = \frac{1}{4}|a-a+0|^2 = 0$$

$$\left|\langle\sigma-|R|\sigma+\rangle\right|^2_{A_{1g}} \propto \frac{1}{4}\left|(1\ -i\ 0)\begin{pmatrix}a & 0 & 0\\ 0 & a & 0\\ 0 & 0 & b\end{pmatrix}\begin{pmatrix}1\\ -i\\ 0\end{pmatrix}\right|^2 = \frac{1}{4}|a-a+0|^2 = 0$$

$$\left|\langle\sigma+|R|\sigma-\rangle\right|^2_{E_{g,1}} \propto \frac{1}{4}\left|(1\ i\ 0)\begin{pmatrix} c & 0 & 0 \\ 0 & -c & d \\ 0 & d & 0 \end{pmatrix}\begin{pmatrix} 1 \\ i \\ 0 \end{pmatrix}\right|^2 = \frac{1}{4}|c+c|^2 = c^2$$

$$\left|\langle\sigma-|R|\sigma+\rangle\right|^2_{E_{g,1}} \propto \frac{1}{4}\left|(1\ -i\ 0)\begin{pmatrix} c & 0 & 0 \\ 0 & -c & d \\ 0 & d & 0 \end{pmatrix}\begin{pmatrix} 1 \\ -i \\ 0 \end{pmatrix}\right|^2 = \frac{1}{4}|c+c|^2 = c^2$$

$$\left|\langle\sigma+|R|\sigma-\rangle\right|^2_{E_{g,2}} \propto \frac{1}{4}\left|(1\ i\ 0)\begin{pmatrix} 0 & -c & -d \\ -c & 0 & 0 \\ -d & 0 & 0 \end{pmatrix}\begin{pmatrix} 1 \\ i \\ 0 \end{pmatrix}\right|^2 = \frac{1}{4}|-ci-ci|^2 = c^2$$

$$\left|\langle\sigma-|R|\sigma+\rangle\right|^2_{E_{g,2}} \propto \frac{1}{4}\left|(1\ -i\ 0)\begin{pmatrix} 0 & -c & -d \\ -c & 0 & 0 \\ -d & 0 & 0 \end{pmatrix}\begin{pmatrix} 1 \\ -i \\ 0 \end{pmatrix}\right|^2 = \frac{1}{4}|ci+ci|^2 = c^2$$

From these results, we observed the breathing ($A_{1g}$) and shear modes ($E_g$) separately. For linearly polarized light, only the shear mode ($\underline{E_g}$ mode) is observed in cross-polarization ($\theta_i + \frac{\pi}{2} = \theta_s$). In contrast, both the breathing ($A_{1g}$) and the shear ($E_g$) modes are observed in parallel polarization ($\theta_i = \theta_s$) with the incident and scattered light represented by $\hat{e}_{i,s} = (\cos\theta_{i,s}\ \sin\theta_{i,s}\ 0)$ in back-scattering geometry.

$$I_{A_{1g}} \propto \left|\langle\hat{e}_s|R|\hat{e}_i\rangle\right|^2 = \left|(\cos\theta_s\ \sin\theta_s\ 0)\begin{pmatrix} a & 0 & 0 \\ 0 & a & 0 \\ 0 & 0 & b \end{pmatrix}\begin{pmatrix} \cos\theta_i \\ \sin\theta_i \\ 0 \end{pmatrix}\right|^2 = a^2|\cos(\theta_i - \theta_s)|^2$$

$$I_{E_{g,1}} \propto \left|\langle\hat{e}_s|R|\hat{e}_i\rangle\right|^2 = \left|(\cos\theta_s\ \sin\theta_s\ 0)\begin{pmatrix} c & 0 & 0 \\ 0 & -c & d \\ 0 & d & 0 \end{pmatrix}\begin{pmatrix} \cos\theta_i \\ \sin\theta_i \\ 0 \end{pmatrix}\right|^2 = c^2|\cos(\theta_i + \theta_s)|^2$$

$$I_{E_{g,2}} \propto \left|\langle\hat{e}_s|R|\hat{e}_i\rangle\right|^2 = \left|(\cos\theta_s\ \sin\theta_s\ 0)\begin{pmatrix} 0 & -c & -d \\ -c & 0 & 0 \\ -d & 0 & 0 \end{pmatrix}\begin{pmatrix} \cos\theta_i \\ \sin\theta_i \\ 0 \end{pmatrix}\right|^2 = c^2|\sin(\theta_i + \theta_s)|^2$$

Since two $E_g$ modes are degenerate, Raman intensity is proportional to the superposed form of $c^2|\cos(\theta_i+\theta_s)|^2+c^2|\sin(\theta_i+\theta_s)|^2=c^2$.

We used the 632.8 nm of a He–Ne laser for Raman measurements. The laser beam was focused onto the sample by a 40× microscope objective lens (0.6 N.A.), and the scattered signal was collected and collimated by the same objective lens. The scattered signal was dispersed with a Jobin-Yvon Horiba iHR550 spectrometer (2400 grooves/mm) and detected with a liquid-nitrogen-cooled charge-coupled-device (CCD) detector. Reflective volume holographic filters (OptiGrate) were used to reject the Rayleigh-scattered light to access the low-frequency range below 100 cm$^{-1}$. The samples were kept in a vacuum chamber during Raman measurements and the laser power was kept below 0.1 mW in order to avoid degradation. For linearly and circularly polarized Raman measurements, $\lambda/2$ and $\lambda/4$ waveplates were used to rotate or convert the polarization of the incident and scattered light. An analyzer selectively passed signals using parallel and cross-polarization configurations.

**Supplementary Note 4. Exponential relaxations in the extended time delay range**

In the time scale of hundreds of picosecond, the exponential relaxations are observed. Supplementary Fig. S7 shows the strain curve of (a) 26 QL and (b) 16 QL. Additional UTXRD experiment for (006) Bragg peak was carried out at 7-ID-C, Advanced Photon Source, USA. The rocking curve measurements are implemented for each laser- x-ray delay position with ±0.8° centered at (006) Bragg peak position. The time resolution of the experimental setup was 50 ps and the minimum time step chosen was 100 ps. The pump laser of 800 nm wavelength was used. The relaxation of expansive strain up to the 8 ns time delay fits with a double-exponential relaxation function. The time constants are classified; one the relaxation time constant $\tau_{rlx}^1$ also observed in UTXRD in XFEL to $\Delta t$~1000ps and the additional relaxation time constant $\tau_{rlx}^2$ dominant in the nanosecond time scale. We found $\tau_{rlx}^2$ dominant after ~1.5 ns.

**Supplementary Figures**

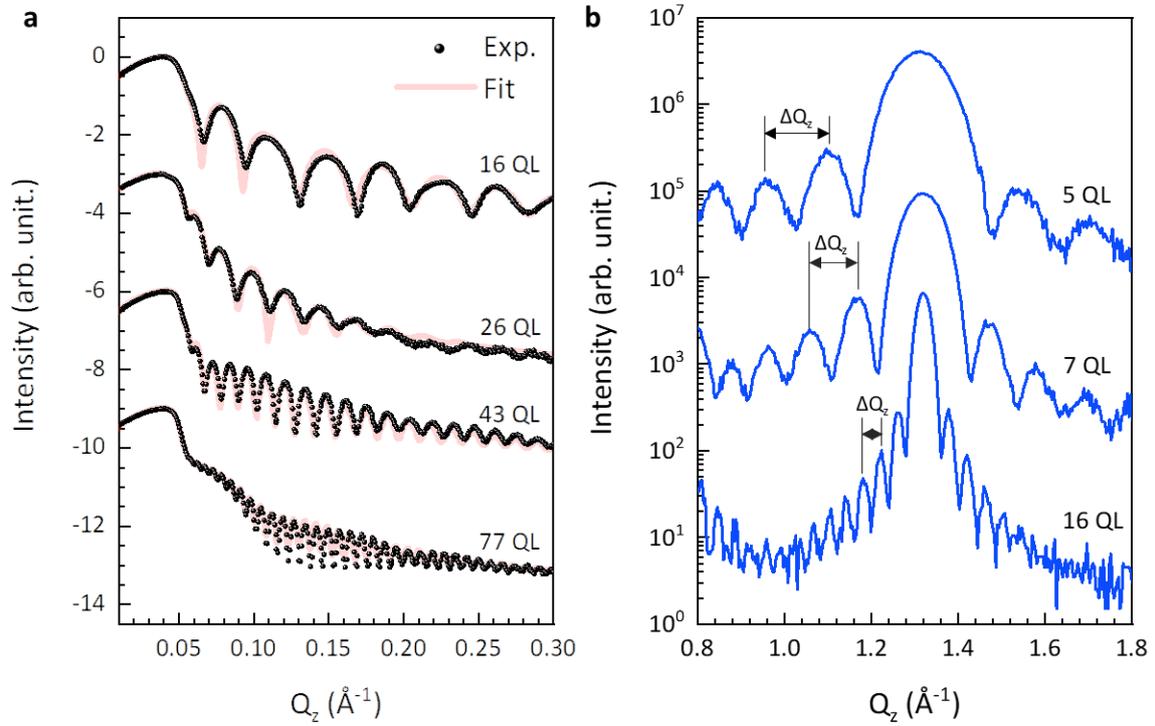

**Supplementary Figure 1. X-ray reflectivity and (006) Bragg X-ray diffraction measurements of the $Bi_2Se_3$ thin films used in the UTXRD.** (a) The film thicknesses of 16, 26, 43, and 77 QL are confirmed by the fitting of X-ray reflectivity intensity data. (b) The thickness of the 5, 7, and 16 QL sample is determined by the fringe gap ($\Delta Q_z$).

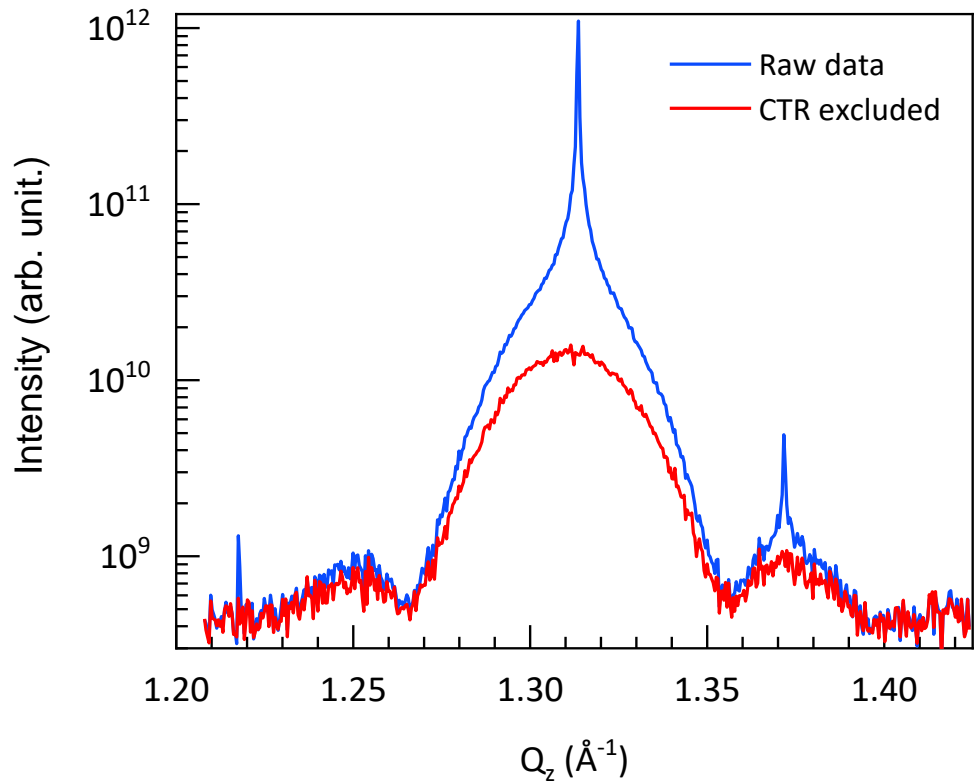

**Supplementary Figure 2. In-plane mean valued intensity data of Figure 1 c in the main text.** Mean valued raw data (red) and CTR excluded data (blue) are displayed. The intensity and the movement of COM of the (006) Bragg peak are determined free from CTR contribution.

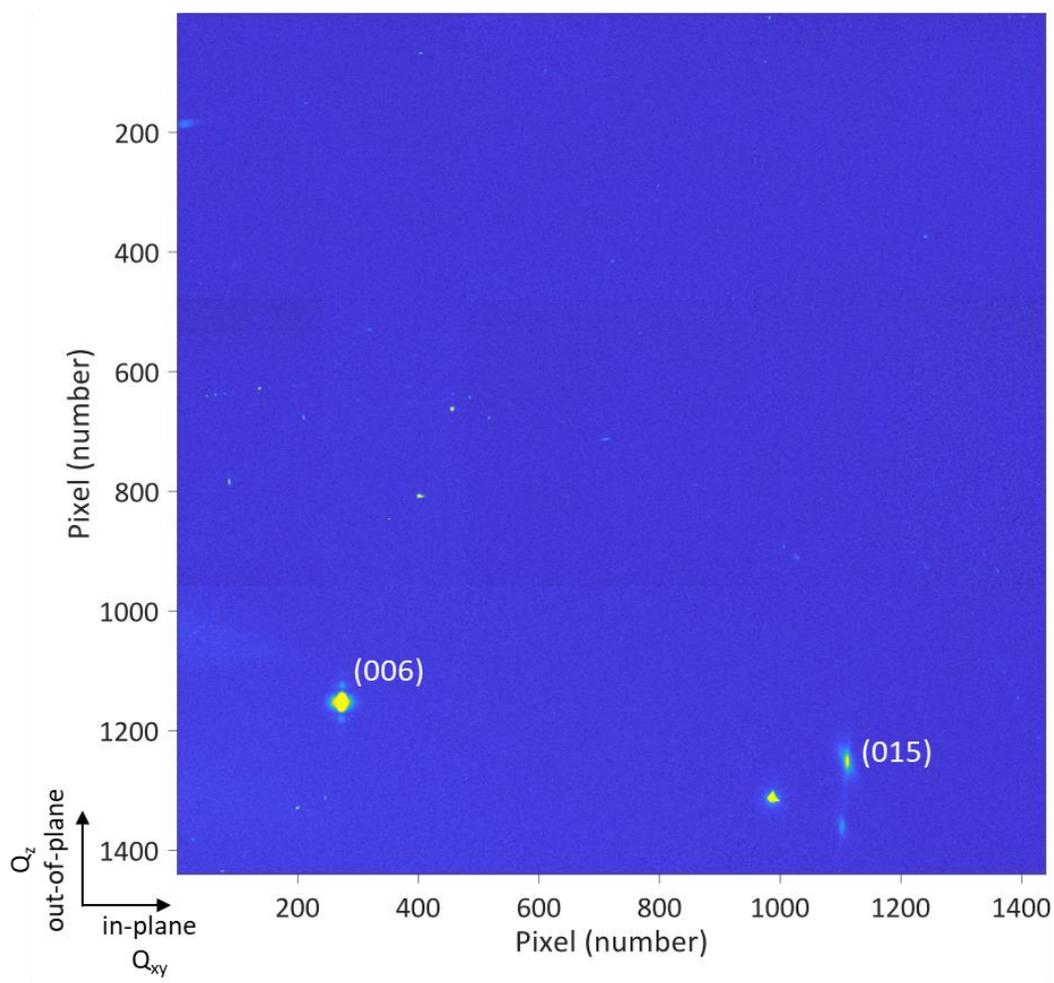

**Supplementary Figure 3. An image of (006) and (015) Bragg peaks under simultaneous diffraction conditions.** (a) (006) and (015) Bragg peaks diffracted on the detector at once. In this geometry, the vertical (horizontal) direction corresponds to $Q_z$ ($Q_{xy}$) and out-of-plane (in-plane) direction.

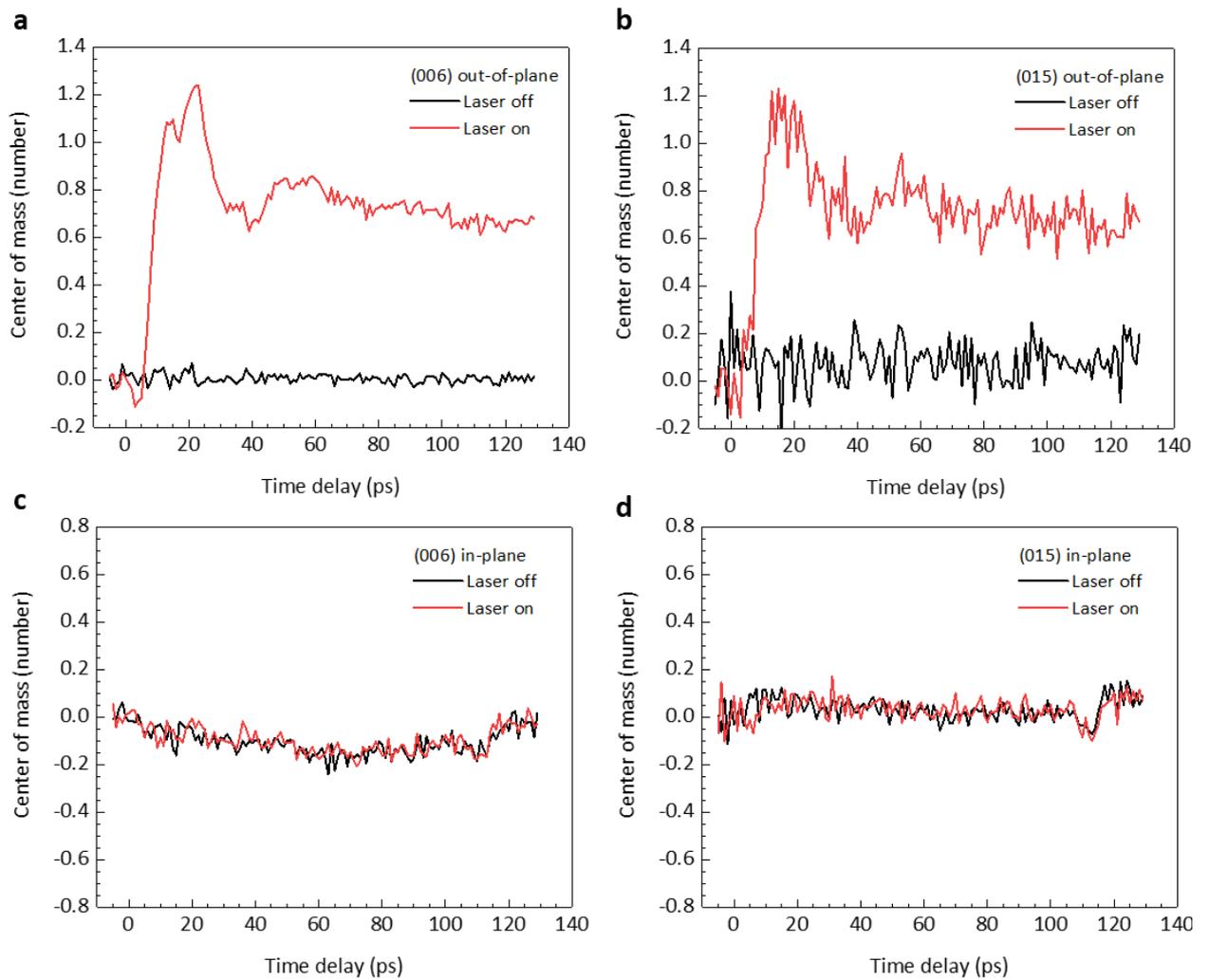

**Supplementary Figure 4. In-plane and out-of-plane change of COM of the (006) and (015) Bragg peaks.** (a-b) Out-of-plane components of COM of (006) Bragg peak with (red) and without laser (black) are shown. (c-d) In-plane components of COM of (006) Bragg peak with (red) and without laser (black) are shown.

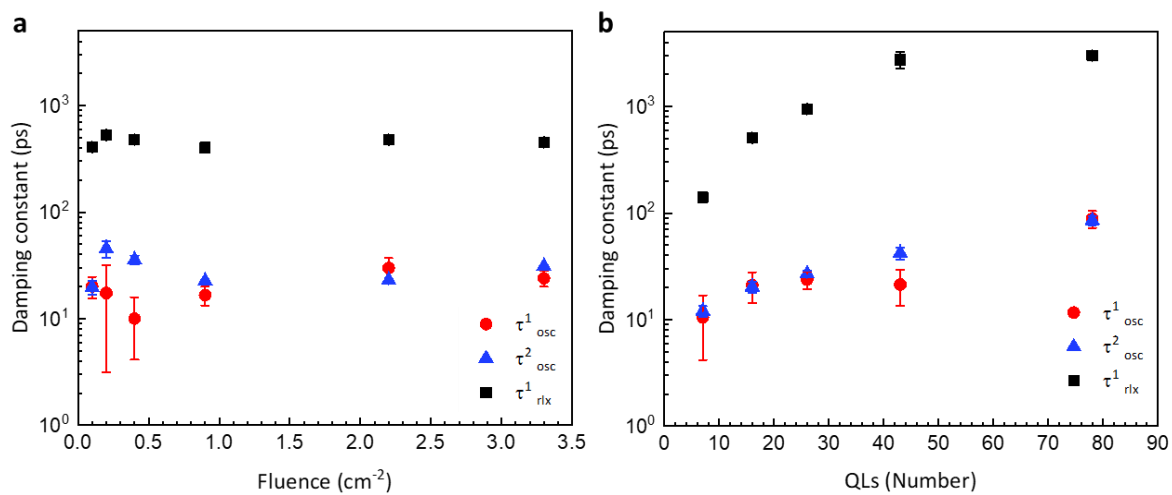

**Supplementary Figure 5. The damping constant from the fit**. The damping constants obtained from the fit are plotted to (a) the fluence for 16 QL and (b) the thickness of the samples with 1.1 mJcm$^{-2}$ of laser fluence. The damping constant is almost no change with fluence, whereas it shows linear dependence with the thickness.

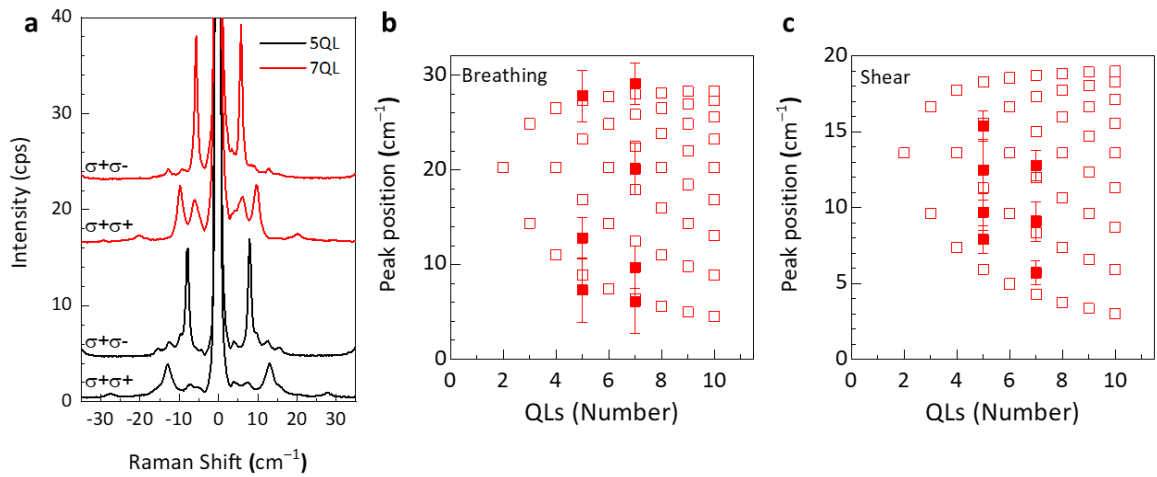

**Supplementary Figure 6. Circularly polarized Raman spectroscopy result.** (a) Circularly polarized Raman spectra measured from 5 and 7 QL samples. Simple linear chain model fit the Raman shifts of (b) breathing mode with $K_z = 5.48 \times 10^{19} \text{Nm}^{-3}$ and (c) shear mode with $K_x = 2.45 \times 10^{19} \text{Nm}^{-3}$.

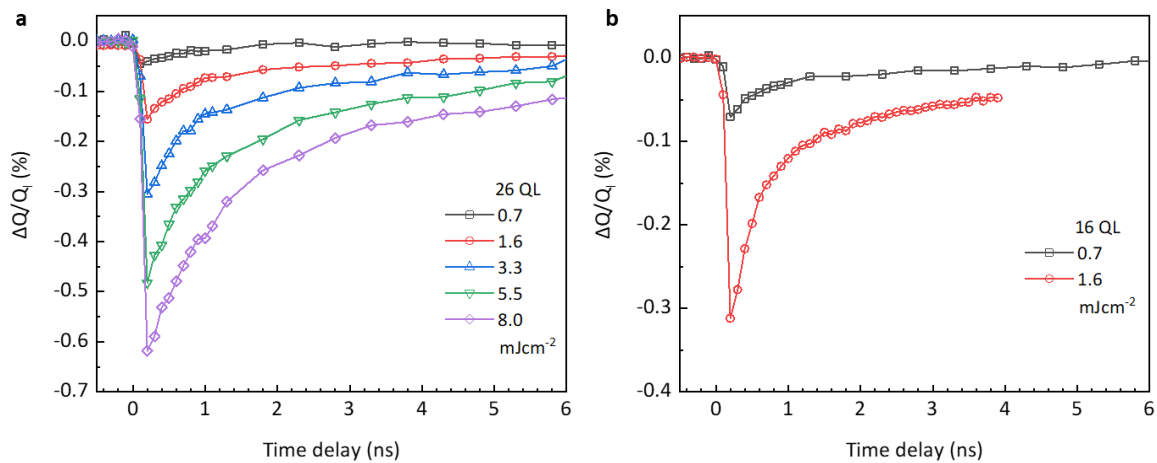

**Supplementary Figure 7. The result of UTXRD in extended time delay range.** (a) Laser-induced strain extracted from (006) Bragg peak of (a) 26QL and (b) 16 QL $Bi_2Se_3$ in nanosecond time scale are shown. Ultrafast time-resolved X-ray diffraction experiment for (006) Bragg peak was carried out at 7-ID-C, Advanced Photon Source, USA. The single data point for each laser- x-ray delay position is obtained from the rocking curve measurement with ±0.8° centered at (006) Bragg peak position. The time resolution of the experimental setup was 50 ps and the minimum time step was 100 ps.

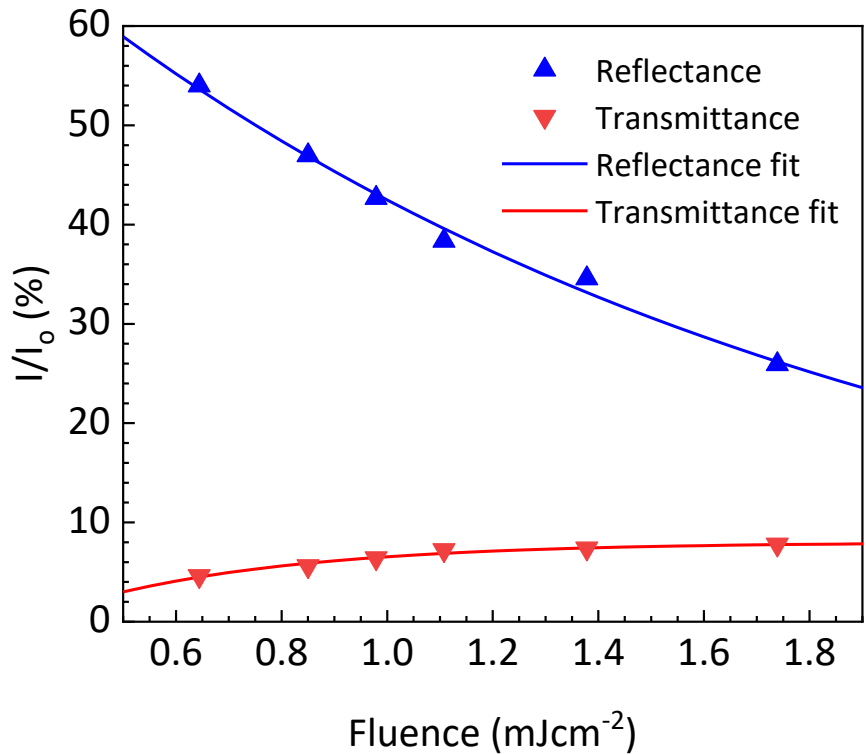

**Supplementary Figure 8. The result of 800 nm laser reflectance and transmittance measurement to the 16 QL sample.** We measured the transmission and the reflectance of the 800 nm laser pulse for the sample used in the UTXRD experiment to get the number of absorbed photons to the sample. The excited carrier density per unit volume ($cm^{-3}$) is obtained by dividing the number of photons per unit area ($cm^{-2}$) by the thickness of the sample. Note that the total thickness was considered for calculating the carrier concentration ($n$) since the penetration depth of the laser is estimated ~81 nm at the incident angle of 15.6° [3].

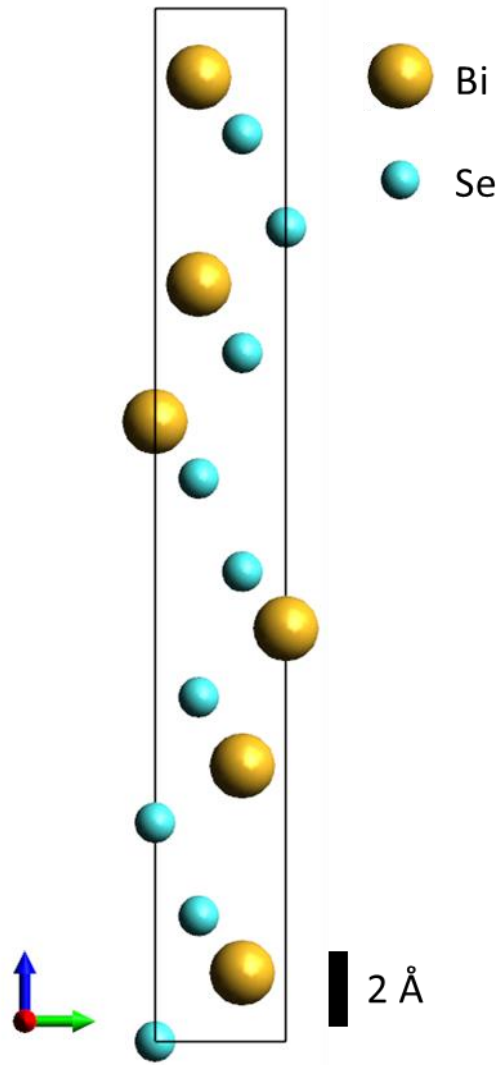

**Supplementary Figure 9.** The unit cell structure of Bi$_2$Se$_3$ used in Density Functional Theory calculations.

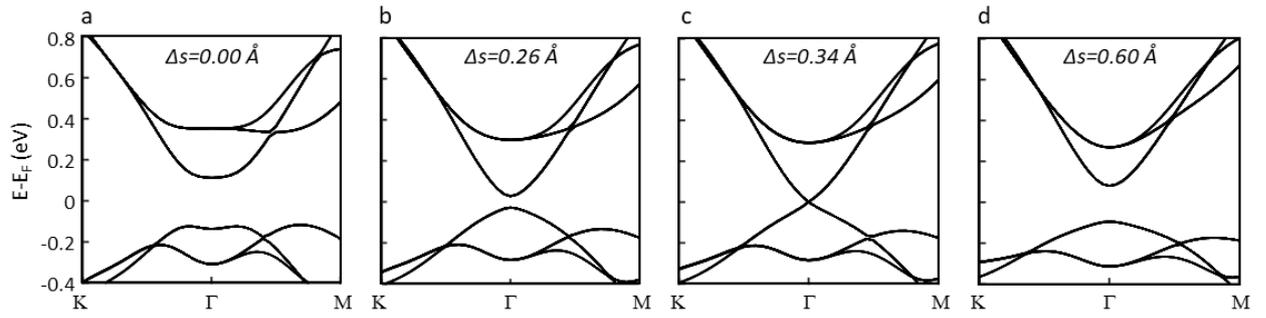

**Supplementary Figure 10. DFT calculated band structure with interlayer distance expansion.** (a) Equilibrium band structure. (b) Band structure with 0.26 Å expansion which is observed maximum expansion on 16 QL. (c) Band structure with 0.34 Å expansion where the band touching is observed. (d) Band structure with 0.6 Å expansion where the topological phase transition to the normal insulator is observed.

**Supplementary Tables**

**Supplementary Table 1. The carrier density was obtained from the fluences used in UTXRD.**

| Fluence (mJcm$^{-2}$) | Reflectance (%) | Transmittance (%) | Absorbance (%) | electron/cm$^3$ ($\times 10^{19}$) |
|---|---|---|---|---|
| 0.02 | 79.00 | 0.00 | 21.00 | 1.06 |
| 0.04 | 79.00 | 0.00 | 21.00 | 2.11 |
| 0.06 | 76.00 | 0.00 | 24.00 | 3.62 |
| 0.1 | 76.00 | 0.00 | 24.00 | 6.04 |
| 0.14 | 74.10 | 0.00 | 25.90 | 9.13 |
| 0.2 | 71.72 | 0.00 | 28.28 | 14.24 |
| 0.4 | 63.00 | 1.60 | 35.40 | 35.64 |
| 1.1 | 39.79 | 6.86 | 53.35 | 147.72 |
| 2.2 | 19.37 | 7.92 | 72.71 | 402.64 |
| 3.3 | 9.426 | 8.00 | 82.58 | 685.98 |

**Supplementary Table 2. Lattice constants of Bi2Se3 unit cell used in DFT.**

| | x (Å) | y (Å) | z (Å) |
|---|---|---|---|
| $a_1$ | 4.17000 | 0.00000 | 0.00000 |
| $a_2$ | -2.08500 | 3.61133 | 0.00000 |
| $a_3$ | 0.00000 | 0.00000 | 28.64000 |

**Supplementary Table 1. Optimized atomic position of Bi$_2$Se$_3$ unit cell.**

|    | x        | y       | z        |
|----|----------|---------|----------|
| Bi | 0.00000  | 2.40755 | 1.90340  |
| Bi | 0.00000  | 2.40755 | 7.64327  |
| Bi | 2.08500  | 3.61133 | 11.45006 |
| Bi | 0.00000  | 0.00000 | 17.18994 |
| Bi | 2.08500  | 1.20378 | 20.99673 |
| Bi | 2.08500  | 1.20378 | 26.73660 |
| Se | 0.00000  | 0.00000 | 0.00000  |
| Se | 2.08500  | 1.20377 | 3.48112  |
| Se | 4.17000  | 0.00000 | 6.06554  |
| Se | 2.08500  | 1.20378 | 9.54667  |
| Se | 0.00000  | 2.40755 | 13.02779 |
| Se | 2.08500  | 1.20378 | 15.61221 |
| Se | 0.00000  | 2.40755 | 19.09333 |
| Se | -2.08500 | 3.61133 | 22.57446 |
| Se | 0.00000  | 2.40755 | 25.15888 |

**Supplementary References**

1. Tolan, M. X-ray Scattering from Soft-Matter Thin Films; Springer, (1999).

2. Parratt, L. G. Surface studies of solids by total reflection of x-rays. *Phys. Rev.* **95**, 359–369 (1954).

3. Kannan, A. G. & Manjulavalli, T. E. Structural, optical and electrical properties of $Bi_2Se_3$ thin films prepared by spray pyrolysis technique. *Int. J. ChemTech Res.* **8**, 599–606 (2015).